\documentclass[twocolumn,aps,prd,10pt]{revtex4-1}

\usepackage{amsmath}
\usepackage{amssymb}
\usepackage{amsfonts}

\newcommand{\gap}{\rule{0pt}{17pt}}

\newcommand{\eref}{\eqref}

\newcommand{\HH}{\mathcal{H}}
\newcommand{\LL}{\mathcal{L}}
\newcommand{\CC}{\mathcal{C}}
\newcommand{\dd}{\text{d}}
\newcommand{\ppp}{\\ \gap}
\newcommand{\nppp}{\nonumber\\ \gap}
\newcommand{\apx}[1]{appendix~\ref{#1}}
\newcommand{\Sec}[1]{Sec.~\ref{#1}}

\newcommand{\Ref}[1]{Ref.~\cite{#1}}
\newcommand{\Refs}[1]{Refs.~\cite{#1}}
\newcommand{\refs}[1]{~\cite{#1}}
\newcommand{\eq}[1]{~\eqref{#1}}

\newcommand{\Eq}[1]{Eq.~\eqref{#1}}
\newcommand{\Eqq}[2]{Eqs.~\eqref{#1} and \eqref{#2}}
\newcommand{\Eqqq}[3]{Eqs.~\eqref{#1}, \eqref{#2}, and \eqref{#3}}

\begin{document}

\title{Immirzi parameter without Immirzi ambiguity: Conformal loop\\ quantization of scalar-tensor gravity}

\author{Olivier J. Veraguth}
\email{o.veraguth@abdn.ac.uk}

\author{Charles H.-T. Wang}
\email{Corresponding author. c.wang@abdn.ac.uk}
\affiliation{Department of Physics,
University of Aberdeen, King's College, Aberdeen AB24 3UE, United Kingdom
\vspace{10 pt}}

\begin{abstract}
Conformal loop quantum gravity provides an approach to loop quantization through an underlying conformal structure i.e. conformally equivalent class of metrics. The property that general relativity itself has no conformal invariance is reinstated with a constrained scalar field setting the physical scale. Conformally equivalent metrics have recently been shown to be amenable to loop quantization including matter coupling. It has been suggested that conformal geometry may provide an extended symmetry to allow a reformulated Immirzi parameter necessary for loop quantization to behave like an arbitrary group parameter that requires no further fixing as its present standard form does. Here, we find that this can be naturally realized via conformal frame transformations in scalar-tensor gravity. Such a theory generally incorporates a dynamical scalar gravitational field and reduces to general relativity when the scalar field becomes a pure gauge. In particular, we introduce a {\it conformal Einstein frame} in which loop quantization is implemented. We then discuss how different Immirzi parameters under this description may be related by conformal frame transformations and yet share the same quantization having, for example, the same area gaps, modulated by the scalar gravitational field.
\\ \vspace{30 pt}
\end{abstract}

\maketitle

\section{Introduction}

Loop quantum gravity (LQG) offers a major nonperturbative approach, through mathematically tractable and conceptually appealing constructions, to quantizing general relativity (GR) that circumvents its nonrenormalizability using conventional quantum field theory\refs{Ashtekar2017, Rovelli2011, Thiemann2008}. Although LQG is not a unified theory {\it per se}, recent demonstrations of its coupling to other fields are important necessary features as a viable theory of quantum gravity. The formulated interactions with Yang-Mills fields do not follow straightforwardly despite the gauge structure inherent in the symmetries of LQG, since LQG has developed its own extensive unique representations in terms of spin networks and spin foams.

Physically indispensable couplings have also been established recently with fermions and most recently with scalar bosons\refs{Ash03, Lewandowski2016}. The coupling to scalar fields in particular has wide implications since they go far beyond merely a form of matter. Scalar fields are responsible for generating mass through the Higgs mechanism, and inducing cosmic inflation as inflatons, may serve to resolve the problem of time\refs{Rovelli1994, Barbour2017}, and provide models for a variety of problems in physics and cosmology\refs{Wang2013, Wang2016, Boj15}.

Leaving matter couplings aside, the incorporation of scalars has made possible the extensions of LQG to beyond-Einstein $f(R)$ and scalar-tensor (ST) theories of gravity\refs{Zhang2011a, Olm11, Zhang2011b}. The purpose of this paper is to address a fundamental issue of LQG---the Immirzi ambiguity---using a constrained or dynamical gravitational scalar field. This ambiguity arises when Ashtekar's original complex ``new variables''\refs{Ash86, Ash87} for an SU$(2)$ spin-gauge connection formalism of GR were revised by Barbero to be real ``Ashtekar-Barbero'' connection variables\refs{Bar95} as a basis for the subsequent LQG\refs{Thiemann2008}. Immirzi was quick to point out that these variables involve a free ``(Barbero-)Immirzi'' parameter $\beta$ and went on to show that a different choice of this  parameter leads to inequivalent quantum theories of gravity having different eigenspectra of operators\refs{Immirzi1997a, Immirzi1997b, Rovelli1998}.

Indeed, the discrete volumes and areas formulated by Rovelli and Smolin\refs{Rovelli1995} depend on the values of $\beta$ entering into LQG, resulting in unitarily unrelated quantizations. While this somewhat unexpected theoretical ambiguity has persisted, to date the mainstream view seems to be taking a ``pragmatic'' approach by fixing $\beta$\refs{Sin14} to phenomenologically match the black hole entropies predicted by the resulting LQG with the Bekenstein-Hawking values. However, some felt such a practice unnatural as exemplified by Rovelli's opinion that ``the result is not entirely satisfactory,'' remarking ``the sense that there is something important which is not yet understood is unavoidable''\refs{Rovelli2011}, while some others have developed models with the Immirzi parameter turned into a scalar field\refs{Taveras2008, Cianfrani2009, Calcagni2009}. Given gradually impending contact of LQG with the real world\refs{Ashtekar2017, Rovelli2011} and increased understanding of quantum gravity effects such as decoherence through the emission and absorption of gravitons\refs{Oniga2016a, Oniga2016b, Oniga2017a, Quinones2017, Oniga2017b, Bassi2017}, the physical ramifications of the Immirzi parameter on spacetime fluctuations towards the deep Planckian domain have been receiving ever more interest and attention\refs{Ashtekar2014}.

At the outset, Immirzi suggested resolving this ambiguity may require ``a group larger than SU(2)''\refs{Immirzi1997b}. Some time ago, motivated by York's conformal analysis of dynamical freedoms of gravity\refs{York1971} and the scaling properties of loop-quantized geometry\refs{Wang2005a}, one of us considered extending the kinematics of LQG to accommodate conformal symmetry\refs{Wang2005b, Wang2006a} leading to the development of ``conformal loop quantum gravity''\refs{Wang2006b} using conformally transformed variables. As GR is not conformally invariant, its conformally extended phase space is subject to a new conformal constraint\refs{Wang2005b, Wang2006a}. Further recent theoretical motivations and justifications for the conformally invariant LQG variables can be found in \Refs{Reid2014, Campiglia2017, Wong2017}.

The purpose of this paper is to report on a new conformal quantization scheme for general ST gravity coupled to matter, containing general relativity as a special case, that is amenable to LQG implementations. By taking advantage of the freedom of changing conformal frames and the conformal invariance of the Einstein gravitational action in the sense of ST theory, we show that when LQG is implemented with a conformally transformed Einstein metric, different values of the corresponding Immirzi parameter are related by a global change of conformal frame. This novel feature suggests the resulting conformal loop quantum ST gravity will be free from the Immirzi ambiguity associated with standard LQG and many of its variants in the literature. In particular, we discuss the prospect of quantized areas with different choices of $\beta$ to have the same discrete spectrum albeit modulated by a power of the scalar field that could arise from the microscopic gravitational constant and new quantum behaviour of geometry at the Planck scale.

We use metric signature $(-,+,+,+)$ with $a,b,\ldots=1,2,3$ and $\alpha,\beta,\ldots=0,1,2,3$ as spatial and spacetime coordinate indices, respectively, and  $i,j,\ldots=1,2,3$ and $I,J,\ldots=0,1,2,3$ as triad and tetrad indices, respectively.

\vspace{10 pt}

\section{Conformal Einstein frame in scalar-tensor gravity}
\label{sec:cef}

Among various potential physical effects of the Immirzi parameter is that it may effectively shift the gravitational constant\refs{Dreyer2003, Jacobson2007}. Combined with ongoing interest in the possible role of conformal properties in LQG including its implications for the Immirzi ambiguity, this consideration leads to the following framework for the loop quantization of ST gravity that is invariant under changes of conformal frames.

Two types of conformal frame for ST gravity have received particular attention in the literature\refs{Wagoner1970}: (a) the Einstein frame in which the gravitational action has a leading term identical to the Hilbert action and (b) the Jordan frame in which the matter action is unaffected by the scalar field. In other words, the Einstein frame is somewhat more gravity oriented with the Jordan frame more matter oriented.

To relate more directly with the standard loop quantization of gravity, we find it useful to start from the Einstein frame as adopted in \Ref{Wang2013} rather than the Jordan frame as adopted in \Refs{Olm11, Zhang2011b}. Furthermore, a global conformal invariance of the gravitational action to be made clear below allows us to formulate a new loop quantization scheme where different Immirzi-type parameters can be conformally related.

%\vspace{20pt}

In the Einstein frame, using the {\it Einstein metric} $\bar{g}_{\alpha\beta}$, with scalar curvature $R[\bar{g}]$, and $\bar{g} = |\det\bar{g}_{\alpha\beta}|$, and the scalar field $\bar{\phi}$, the total Lagrangian (density) for a general ST gravity is given by \Ref{Wang2013} to be
\begin{eqnarray}
\LL
&=&
\LL_\text{ES}+\LL_\text{SP}+\LL_\text{M},
\label{LL}
\gap
\end{eqnarray}
where
\begin{eqnarray}
\LL_\text{ES}
&=&
\frac1{2\kappa}
\Big[
\sqrt{\bar{g}}R[\bar{g}]
-
2\sqrt{\bar{g}}\bar{g}^{\alpha\beta}\bar{\phi}_{,\alpha}\bar{\phi}_{,\beta}
\Big]
\label{LE}
\gap
\end{eqnarray}
is here referred to as the Einstein-scalar Lagrangian,
\begin{eqnarray}
\LL_\text{SP}
&=&
-
\frac2\kappa\sqrt{\bar{g}}\, V(\bar{\phi})
\label{LP}
\gap
\end{eqnarray}
is a scalar potential Lagrangian for some potential function $V(\bar{\phi})$, and
\begin{eqnarray}
\LL_\text{M}
&=&
\LL_\text{M}[\Omega^2(\bar{\phi})\bar{g}_{\alpha\beta},\psi]
\label{LM}
\gap
\end{eqnarray}
is a matter Lagrangian for some metric coupling function
$\Omega(\bar{\phi})$ and matter fields $\psi$.
In the above, $\kappa$ is a coupling constant which may be identified as
$\kappa=8\pi {G}/c^4$ if the ST theory reduces to GR with $\bar{\phi}=$ const. For the Brans-Dicke theory,
$\Omega^2(\bar\phi) \propto \exp[-\bar{\phi}/\sqrt{\omega+3/2}]$
using the Brans-Dicke coupling constant $\omega$. GR is also obtained in the limits ${\omega\to-3/2}$ and ${\Omega\to}$ const.
The combination $\Omega(\bar{\phi})\bar{g}_{\alpha\beta}$ is called the ``physical metric''\refs{Wagoner1970} as it defines the spacetime geometry that matter ``feels.''

%\vspace{20pt}

Below we will focus on the Einstein-scalar Lagrangian $\LL_\text{ES}$ and use it to derive a set of conformal spin-gauge variables for ST gravity. Starting from the scalar-tensor variables
${(\bar{g}_{\alpha\beta}, \bar{\phi})}$
in the Einstein frame, one can can always change to an alternative conformal frame with variables
${({g}_{\alpha\beta}, {\phi})}$
by reparametrizing the scalar field $\bar{\phi}=\bar{\phi}(\phi)$ and conformally transforming the metric tensor
${\bar{g}_{\alpha\beta} = F^2(\phi){g}_{\alpha\beta}}$
for some function $F(\phi)$.
Analogous to previous works on conformal loop quantum gravity\refs{Wang2005b, Wang2006a, Wang2006b, Campiglia2017} with a constrained scalar field $\phi$, we choose $F(\phi)$ so that ${g}_{\alpha\beta}$ is given by
\begin{eqnarray}
\bar{g}_{\alpha\beta}
&=&
\phi^{2}g_{\alpha\beta} .
\label{cnf1}
\gap
\end{eqnarray}
In addition, we adopt the reparametrization
\begin{eqnarray}
\bar{\phi}
&=&
\ln\phi
\label{chiphi}
\gap
\end{eqnarray}
of the scalar field, which is dynamical, except for GR reduction, so that the Einstein-scalar Lagrangian\eq{LE} becomes
\begin{eqnarray}
\LL_\text{ES}
&=&
\frac1{2\kappa}
\Big[
\phi^4\sqrt{g}R[\phi^2g_{\alpha\beta}]
-
2\sqrt{g}g^{\alpha\beta}\phi_{,\alpha}\phi_{,\beta}
\Big] .
\label{LST2}
\gap
\end{eqnarray}

We will refer to the conformal frame with variables ${({g}_{\alpha\beta}, {\phi})}$ obtained from the Einstein frame variables
${(\bar{g}_{\alpha\beta}, \bar{\phi})}$ through \Eqq{cnf1}{chiphi}
as the {\it conformal Einstein frame}. Its usefulness in addressing the Immirzi ambiguity follows from the form of Lagrangian\eq{LST2} being invariant under the following global conformal transformations:
\begin{eqnarray}
\phi
&\rightarrow&
\Lambda^{-1}\phi
%\label{phiconf}
%\\
,\;
g_{\alpha\beta}
\rightarrow
\Lambda^2 g_{\alpha\beta}
\label{Omg}
\gap
\end{eqnarray}
for any positive constant $\Lambda$.

As will become clear later, unlike standard approaches, our idea is to introduce the Ashtekar-Barbero type variables in the conformal Einstein frame and show that any choice of the corresponding Immirzi parameter can be mapped to a different value, such as unity, in an alternative conformal Einstein frame transformed using \Eq{Omg}.

\vspace{10pt}

\section{Canonical analysis in the conformal Einstein frame}
\label{sec:cana}

Prior to canonical quantization of ST gravity in a conformal Einstein frame, the preceding Lagrangian formalism will in this section be transcribed to the corresponding Arnowitt-Deser-Misner canonical formalism. First, using \Eq{cnf1}, we can map the spatial metric $\bar{h}_{ab}$, lapse function $\bar{N}$, and shift vector $\bar{N}^a$ associated with the Einstein metric $\bar{g}_{\alpha\beta}$ to their counterparts  $h_{ab}$, $N$, and $N^a$ according to
\begin{eqnarray}
\bar{h}_{ab}
&=&
\phi^{2}h_{ab}
\label{cnf2}
\ppp
\bar{N}
&=&
\phi N
\label{cnf3}
\ppp
\bar{N}^a
&=&
N^a .
\label{cnf4}
\end{eqnarray}
We then evaluate the extrinsic curvature in the conformal Einstein frame using the above to get
\begin{eqnarray}
K_{ab}
=
\frac1{2N}
\left(-h_{ab,0} + N_{a;b} + N_{b;a}\right) .
\label{defKij}
\gap
\end{eqnarray}
After some calculations involving conformal change relations\refs{Yan70},
the above then yields the Einstein-scalar Lagrangian up to a total divergence of the form
\begin{eqnarray}
\LL_\text{ES}
&=&
\frac{\phi^2}{2\kappa}\sqrt{h}N
\left(K_{ab}K^{ab}-K^2+R[h]\right)
+
\frac2\kappa\sqrt{h}\phi\phi_{,0}K
\nppp&&\hspace{-5pt}
-
\frac{2N^a}\kappa\sqrt{h}\phi\phi_{,a}K
-
\frac{2N}\kappa\sqrt{h}(\phi\phi_{,a})^{;a}
-\frac{2}{\kappa N}\sqrt{h}
\nppp&&\hspace{-5pt}
\times
\Big[
\phi_{,0}^2
-2N^a\phi_{,0}\phi_{,a}
-\left(N^2h^{ab}-N^aN^b\right)\phi_{,a}\phi_{,b}
\Big] .
\nppp
\label{eqL}
\gap
\end{eqnarray}
This gives rise to the canonical momenta for the metric
${p^{ab}={\delta\int\LL_\text{ES}\,\dd^3x}/{\delta h_{ab,0}}}$
and scalar
${\pi_\phi={\delta\int\LL_\text{ES}\,\dd^3x}/{\delta\phi_{,0}}}$
as follows
\begin{eqnarray}
p^{ab}
&=&
-
\frac1{\kappa N}\sqrt{h}\,h^{ab}\phi(\phi_{,0}-\phi_{,c}N^c)
\nppp&&
-
\frac{\phi^2}{2\kappa}\sqrt{h}\,(K^{ab}-h^{ab}K)
\label{eq43}
\gap
\end{eqnarray}
where $h=\det h_{ab}$, and
\begin{eqnarray}
\pi_\phi
=
\sqrt{h}\,\Big[
\frac2\kappa \phi K - \frac4{\kappa N}(\phi_{,0}-\phi_{,c}N^c)
\Big],
\label{eq45}
\gap
\end{eqnarray}
yielding
\begin{eqnarray}
\phi_{,0}
=
\phi_{,c}N^c
-
\frac{\kappa N}{4\sqrt{h}}\pi_\phi + \frac{N}{2} \phi K .
\label{eqphid}
\gap
\end{eqnarray}
It also follows from \Eq{defKij} that
\begin{eqnarray}
h_{ab,0}
&=&
\frac{4\kappa N}{\phi^2\sqrt{h}}
\Big[
p_{ab}
-
\frac14
h_{ab}\phi\pi_\phi
\Big] + N_{a;b} + N_{b;a}
\label{hij0a}
\gap
\end{eqnarray}
By using Eqs.~\eqref{defKij}, \eqref{eqL}, \eqref{eqphid}, and \eqref{hij0a},  up to a total divergence we can derive
\begin{eqnarray}
\HH_\text{ES}
&=&
p^{ab}h_{ab,0}+\pi_\phi\phi_{,0}-\LL_\text{ES}
\nppp
&=&
N\CC_\perp + N^a\CC_a
\label{Hamden}
\gap
\end{eqnarray}
with the Hamiltonian and diffeomorphism constraints as follows:
\begin{eqnarray}
\gap
\CC_\perp
&=&
\frac{2\kappa}{\sqrt{h}}\phi^{-2}
p_{ab}p^{ab}
-
\frac{\phi^2}{2\kappa}\sqrt{h} R[h]
+
\frac{2}\kappa\sqrt{h}h^{ab}\phi\phi_{,a;b}
\nppp&&
+
\frac{\kappa}{4{\sqrt{h}}}
\,{\pi_\phi^2}
-
\frac{\kappa}{{\sqrt{h}}}
\,{\phi^{-1}\pi_\phi p}
\label{CC2}
\ppp
\CC_a
&=&
-2  h_{ab}p^{bc}{}_{;c}
+\phi_{,a}\pi_\phi
\label{CCi2}
\gap
\end{eqnarray}
in terms of $p = h_{ab}p^{ab}$.

From the totally constrained nature of the Hamiltonian\eq{Hamden}, $\CC_\perp$ and $\CC_a$ are required to vanish weakly: $\CC_\perp \approx 0, \CC_a \approx 0$.
The consistency of this is ensured by the Dirac algebra of $\CC_\perp$ and $\CC_a$ under their Poisson brackets as established in \apx{app:dirac}, which is  satisfied also by the Hamiltonian and diffeomorphism constraints for canonical GR.

The canonical generator of the local conformal transformation that preserves the Einstein metric $\bar{g}_{\alpha\beta}$ is given by
\begin{eqnarray}
%\CC_\angle
\CC
&=&
%\phi\pi_\phi -2h_{ab}p^f{ab}
%=
\phi\pi_\phi -2p .
\label{cf}
\gap
\end{eqnarray}
For the GR case, we require in addition the weakly vanishing
$\CC\approx0$
as well, with the consistent closure condition under Poisson brackets previously established in \Refs{Wang2005b, Wang2006a}. In general, the uniformed smeared $\CC$ generates the discussed global conformal invariance of $\LL_\text{ES}$ using \Eq{Omg} as will be discussed further below.

%\section{Conformal spin-gauge variables for scalar-tensor gravity}
%\vspace{25pt}

\section{Conformal Ashtekar-Barbero variables and conformal Immirzi parameter}
\label{sec:cab}

%\section{Canonical transformation to the triad variables}

Having obtained the Hamiltonian formalism of ST gravity in the conformal Einstein frame with variables ${({g}_{\alpha\beta}, {\phi})}$, we can now proceed to finding their counterpart for the standard Ashtekar-Barbero variables and explore the property of the  resulting Immirzi parameter in what follows.

%\vspace{20pt}

In terms of the triad $e^a_i$ and its densitization
${E^a_i=\sqrt{h}\, e^a_i}$ associated with $h_{ab}$, we have the standard relation
\begin{eqnarray}
h_{ab}
&=&
\delta_{ij} e^i_a e^j_b
=
h\,\delta_{ij} E^i_a E^j_b .
\label{hab}
\gap
\end{eqnarray}
Substituting \eqref{eqphid} into \eqref{eq43} we obtain
\begin{eqnarray}
p^{ab}
&=&
-
\frac12 \CC h^{ab}
-
\frac{\phi^2}{2\kappa}\sqrt{h}(K^{ab}-h^{ab}K) .
\label{pab}
\gap
\end{eqnarray}
It then follows from \Eq{hab} that
\begin{eqnarray}
h_{ab,0}
&=&
\frac1h (h_{ab} h_{cd} -h_{ac}h_{bd}-h_{ad}h_{bc})E^c_i E^d_{i,0}
\label{hab0a}
\gap
\end{eqnarray}
and hence
\begin{eqnarray*}
p^{ab}h_{ab,0}
&=&
E^a_i K^i_{a,0}
- (E^a_i K^i_a)_{,0}
\gap
\end{eqnarray*}
where
\begin{eqnarray}
K^i_a
&=&
-
\frac{1}{\kappa\sqrt{h}} \phi^2K_{ab}E^b_i
+
\frac{1}{2h} \CC h_{ab} E^b_i .
\label{Kia}
\gap
\end{eqnarray}
Contracting \eqref{Kia} with $E^i_c$ we get
\begin{eqnarray}
K_{ab}
&=&
-
\kappa\sqrt{h}\phi^{-2}
K^i_a E^i_b
+
\frac{\kappa}{2\sqrt{h}}
\phi^{-2} \CC h_{ab} .
\label{Kia1}
\gap
\end{eqnarray}
Using $K_{ab}=K_{ba}$ we see from \Eq{Kia} that $K^i_a E^i_b = K^i_b E^i_a$.
However, we will from now on treat $K^i_a$ and $E^a_i$ as independent variables without imposing this condition.
Instead we define
\begin{eqnarray}
K_{ab}
&=&
-
\kappa\sqrt{h}\phi^{-2}
K^i_{(a} E^i_{b)}
+
\frac{\kappa}{2\sqrt{h}}
\phi^{-2} \CC h_{ab}
\label{defKia}
\gap
\end{eqnarray}
in terms of arbitrary $K^i_a$ and $E^a_i$.
Then by using \Eqq{pab}{defKia} we can calculate that
\begin{eqnarray}
p^{ab}
&=&
\frac{h}{4}
(h^{ad}h^{bc}+h^{ac}h^{bd}-2h^{ab}h^{cd})
K^i_{c} E^i_{d} .
\label{pabKE}
\gap
\end{eqnarray}
By contracting the above with $h_{ab}$, we have
\begin{eqnarray}
p
&=&
-{h}h^{ab}K^i_{a} E^i_{b}
=
-K^i_{a} E^a_{i} .
\label{pKE}
%\gap
\end{eqnarray}
Then using \eqref{pabKE} and \eqref{hab0a} we can evaluate that
\begin{eqnarray}
p^{ab}h_{ab,0}
&=&
-
K^j_{b} E^b_{j,0}
-
K^i_{[a} E^i_{b]} E^a_j E^b_{j,0} .
\gap
\end{eqnarray}
This implies an additional constraint $K^i_{[a} E^i_{b]}$ or equivalently
\begin{eqnarray}
\CC_k = \epsilon_{kij}K_{a[i}E^a_{j]} = - \epsilon_{kij} K^l_{[a} E^l_{b]} E^a_i E^b_j
\label{spincons}
\gap
\end{eqnarray}
called the spin constraint\refs{Wang2005b}.
It is equivalent to the rotation constraint defined in \Ref{Thiemann2008} and generates local SU(2) transformations\refs{Henneaux1989}.
From \Refs{Wang2005b, Zhang2011b}, this constraint is first class, forming a closed Poisson bracket algebra with the Hamiltonian, diffeomorphism, and conformal constraints.

The following variables then form canonical pairs:
\begin{eqnarray}
(\kappa K^i_a, \kappa^{-1}E^a_i) \text{ and } (\phi, \pi_\phi) .
\label{cano}
\end{eqnarray}
Using these canonical variables and \eqref{pKE}, we see that the conformal constraint for GR\eq{cf} becomes
\begin{eqnarray}
\CC
&=&
\phi\pi_\phi + 2 K^i_{a} E^a_{i} .
\label{cf2}
\gap
\end{eqnarray}
%\section{Canonical transformation to the connection variables}
For any positive constant $\beta$, a trivial canonical transformation from \eqref{cano} yields
\begin{eqnarray}
(\beta \kappa K^i_a, \beta^{-1}\kappa^{-1}E^a_i) \text{ and } (\phi, \pi_\phi) .
\label{cano0}
\gap
\end{eqnarray}
Since $\Gamma^i_a$ commute with $E^a_i$ under Poisson brackets, we can further perform a canonical transformation from \eqref{cano} to yield
\begin{eqnarray}
(A^i_a = \Gamma^i_a+\kappa K^i_a, \kappa^{-1}E^a_i) \text{ and } (\phi, \pi_\phi)
\label{cano1}
\end{eqnarray}
and alternatively
from \eqref{cano0} to yield
\begin{eqnarray}
(A'^i_a = \Gamma^i_a+\beta\kappa K^i_a, \beta^{-1}\kappa^{-1}E^a_i) \text{ and } (\phi, \pi_\phi) .
\label{cano2}
\end{eqnarray}

%\vspace{20pt}

We refer to canonical variables\eq{cano1} and more generally\eq{cano2} as {\it conformal Ashtekar-Barbero} variables and $\beta$ involved as the {\it conformal Immirzi parameter}. For any $\beta$, the variable $A^i_a$ has the same construction as the SU(2) spin connection with densitized triad $E^a_i$ as the conjugate momentum. As they have the same structure as the standard Ashtekar-Barbero variables of LQG, they are amenable to loop quantization based on a spin-network representation with a Hilbert space here denoted by $\HH_\text{SN}$\refs{Ashtekar2017, Rovelli2011, Thiemann2008}.

%\vspace{20pt}

To quantize the scalar-spin variables\eq{cano2} as a whole, we follow the kinematic quantization recently developed in \Ref{Lewandowski2016}, where a diffeomorphism invariant representation using a Hilbert space $\HH_\text{SF}$ for the scalar field in which the field operator $\phi$ is diagonal. This leads to the total Hilbert space
\begin{eqnarray}
\HH=\HH_\text{SF}\otimes\HH_\text{SN}
\end{eqnarray}
as with the treatment of LQG coupled to a scalar field in \Ref{Lewandowski2016}. The quantum states are therefore expressed as superpositions of
\begin{eqnarray}
\Psi[\phi,A]
=
\Psi[\phi]\otimes\Psi[A]
\gap
\end{eqnarray}
in terms of the cylindrical functions of $A$,
\begin{eqnarray}
\Psi[A]
=
\psi(h_{e_1}[A],\ldots,h_{e_n}[A])
\gap
\end{eqnarray}
involving holonomies $h_{e_1}[A],\ldots,h_{e_n}[A]$ over edges
$e_1,\ldots,e_n$.

In terms of the variables $(A^i_a, E^a_i)$, the spin constraint\eq{spincons} can be rewritten as the familiar Gauss constraint
\begin{eqnarray}
\CC_k
&=&
D_a E^a_k
=
E^a_{k,a} +
\epsilon_{kij}A^i_a E^a_j
\label{gausscons}
\gap
\end{eqnarray}
that generates rotations. Since the scalar states $\Psi[\phi]$ are SU(2) invariant and by construction $\Psi[A]$ satisfy the quantum Gauss law, the spectra of invariant operators such as areas on $\Psi[\phi,A]$ are unchanged under actions from \Eq{gausscons}. As with \Eq{spincons}, from \Refs{Wang2005b, Zhang2011b}, the above Gauss constraint is also first class, being closed under Poisson bracket algebra with the Hamiltonian, diffeomorphism, and conformal constraints\refs{Wang2005b, Zhang2011b}.

%\vspace{20pt}

The new conformal properties of the conformal Ashtekar-Barbero variables mean that the standard Immirzi ambiguity no longer has a direct analogy as will be explained below.
By using
\begin{eqnarray}
\left\{
\begin{array}{rl}
\phi'
=
\beta^{1/2}\phi\;,
&
\pi'_\phi
=
\beta^{-1/2}\pi_\phi
\ppp
E'^a_i
=
\beta^{-1} E^a_i,
&
K'^i_a
=
\beta K^i_a
\end{array}
\right.
\label{trans}
\gap
\end{eqnarray}
and the structures of $\CC_\perp$ and $\CC_a$, and $\CC_k$ (see \apx{apx:conf} for details), the canonical variables\eq{cano2} are equivalent to
\begin{eqnarray}
(A'^i_a = \Gamma^i_a+\kappa K'^i_a, \kappa^{-1}E'^a_i) \text{ and } (\phi', \pi'_\phi) .
\label{cano3}
\gap
\end{eqnarray}
With a global conformal transformation,  the canonical variables\eq{cano3} are in turn equivalent to \Eq{cano1}.
Therefore the two sets of canonical variables \Eqq{cano1}{cano2} are equivalent.

At the quantum level the global conformal transformation of a quantum state $\Psi[\phi,A]$ is generated by the uniformly smeared conformal constraint
\begin{eqnarray}
C = \int\CC\,\dd^3x
\gap
\end{eqnarray}
as follows:
\begin{eqnarray}
\Big(1-\frac{i\epsilon C}{\hbar}\Big)
\Psi[\phi,A^a_i]
=
\Psi[\phi+\epsilon\phi, A^a_i + 2\epsilon \kappa K^a_i]
\gap
\end{eqnarray}
for an infinitesimal $\epsilon$,
where $\CC$ given by \Eq{cf2} can be implemented through Thiemann's quantization of ${K^i_{a} E^a_{i}}$ terms\refs{Thiemann2008, Thiemann1996} and Lewandowski and Sahlmann's quantization of
${\pi_\phi\to -i\hbar\,\delta/\delta\phi}$ terms\refs{Lewandowski2016}.

%\vspace{20pt}

The quantum implementation of the invariance under global conformal transformation\eq{trans} causes no ordering issues, as it contributes only to commuting powers of $\beta$ as a $c$ number in e.g. the Hamiltonian constraint $\CC_\perp$ leaving an overall transformation according to \Eq{CCcnf} as
\begin{eqnarray}
\CC'_\perp
&=&
\beta^{1/2} \CC_\perp .
\label{CCtrans}
\gap
\end{eqnarray}

%\vspace{20pt}

Therefore Dirac quantization using $\CC_\perp$ or $\CC'_\perp$ yields the same physics irrespective of the choice of $\beta$. Specifically, we have
invariant discrete areas and volumes using the Einstein frame densitized triad under the global conformal transformation
\begin{eqnarray}
\bar{E}^a_i = \phi^2 {E}^a_i = \phi'^2 {E}'^a_i
%\label{}
\gap
\end{eqnarray}
by using \Eq{trans}, which is clearly independent of the value of conformal Immirzi parameter $\beta$.

%To demonstrate that the Immirzi parameter $\beta$ can be absorbed by a conformal transformation, note that the quantization based on \eqref{cano2} gives quantized areas
%$A \sim \phi^2\beta \sim \phi'^2$
%which is the same as quantization based on \eqref{cano1}.

\vspace{15pt}

\section{Conclusion and outlook}

In this paper we address loop quantization in a wider context of ST gravity. Our main motivation has not been just to extend LQG beyond GR, but to seize on the freedom of conformal frame transformations available in ST gravity that may help rendering the ambiguous Immirzi parameter in current LQG into a more natural conformal gauge parameter having no preferred values. For this purpose we have found it useful to start from the Einstein frame of ST gravity followed by certain conformal frame transformation into the conformal Einstein frame in \Sec{sec:cef}, since the resulting scalar-gravitational action specified by the Einstein-scalar Lagrangian\eq{LE} used to define canonical variables of gravity has a global conformal symmetry\eq{Omg}. This indicates that loop gravitational variables built in such a conformal frame may inherit a global conformal symmetry under which the corresponding Immirzi parameter is the gauge parameter.

We have therefore been led by the above observation to the construction of the Hamiltonian formalism of ST gravity in the conformal Einstein frame in \Sec{sec:cana} as a prerequisite for canonical quantization. Like canonical GR, the resulting Hamiltonian ST system is also totally constrained, having a more involved set of the Hamiltonian constraint\eq{CC2} and diffeomorphism constraints\eq{CCi2}. Furthermore, these constraints satisfy the same Dirac algebra under their Poisson brackets with respect to the conformal Einstein frame canonical variables as explicitly established in \apx{app:dirac}. We then build on this canonical structure a new set of conformal Ashtekar-Barbero variables with a corresponding conformal version of the Immirzi parameter $\beta$ in \Sec{sec:cab}. Remarkably, this conformal Immirzi parameter does indeed represent the global conformal gauge parameter whose value should not appear in physical observables such as area operators after quantization. These main findings also apply to GR as a special case of ST gravity where the conformal constraint equation $\CC\approx 0$ must be satisfied making the scalar gravitational field a pure gauge.

Additionally, we remark that the above canonical treatment may also be approached using a conformal version of the Holst action\refs{Holst1996}, relevant for the spin foam extension to this work. The new starting point would be to consider the Hilbert part of Lagrangian\eq{LE} in its Palatini form
\begin{eqnarray}
\LL_\text{P}
&=&
\frac1{2\kappa}
%\Big[
\bar{e}\,\bar{e}^\alpha{}_I \bar{e}^\beta{}_J \overline{F}_{\alpha\beta}{}^{IJ}
%-
%2\,\bar{e}\,\bar{e}^\alpha{}_I \bar{e}^\beta{}^I
%\bar{\phi}_{,\alpha}\bar{\phi}_{,\beta}
%\Big]
\label{EE}
%\gap
\end{eqnarray}
in terms of the tetrad $\bar{e}^\alpha{}_I$ with determinant $\bar{e}$ and curvature $\overline{F}_{\alpha\beta}{}^{IJ}$ associated with
$\bar{g}_{\alpha\beta}=\phi^{2}g_{\alpha\beta}$ as in \Eq{cnf1}.
A plausible candidate for the conformally modified {\it Palatini-Holst Lagrangian} of \Eq{EE} would then be
%http://solenodonus.com/file/hamiltonian-analysis-of-holst-action.html
\begin{eqnarray}
\LL_\text{PH}
&=&
\frac1{2\kappa}
\Big[
\bar{e}\,\bar{e}^\alpha{}_I\, \bar{e}^\beta{}_J\,
\overline{F}_{\alpha\beta}{}^{IJ}
\nppp&&
-
\frac{1}{2}\,
\underline{e}\,\underline{e}^\alpha{}_I\, \underline{e}^\beta{}_J\,
\epsilon^{IJ}{}_{KL} \underline{F}_{\alpha\beta}{}^{KL}
\Big]
\label{EH}
%\gap
\end{eqnarray}
in terms of the tetrad $\underline{e}^\alpha{}_I$ with determinant $\underline{e}$ and curvature $\underline{F}_{\alpha\beta}{}^{IJ}$ associated with $\underline{g}_{\alpha\beta}=\theta^{2}g_{\alpha\beta}$ using another (multiplier) scalar field $\theta$.

Note that in \Eq{EH}, the extra Holst term is associated with a conformally transformed metric $\underline{g}_{ab}$, nonidentical to $\bar{g}_{ab}$ as used in \Eq{EE} and so this action is different from that in \Ref{Campiglia2017}, which is recovered from \Eq{EH} by equating $\phi$ and $\theta$. On the other hand, the standard Palatini-Holst action is recovered from \Eq{EH} through conformal gauge fixing with constants $\phi=1$ and $\theta=\beta^{-1/2}$ yielding the usual Immirzi parameter $\beta$.
Furthermore, Lagrangian\eq{EH} can be shown to reduce to the Einstein-Hilbert action for GR after varying the scalars $\phi$ and $\theta$ and the connection 1-forms of the tetrad ${e}^\alpha{}_I$. However, leaving $\phi$ and $\theta$ free retains the freedom of conformal frame transformation in the quantum dynamics to be explored in the context of the Immirzi ambiguity.

Finally, although a fully quantum description of the conformal LQG formalism without a free Immirzi parameter proposed in this work is yet to be completed, one may already wonder how the well-accepted Bekenstein-Hawking entropy of black holes could be recovered. Admittedly, here we see little immediate phenomenological analogy of matching the Immirzi parameter as done in standard LQG. Nonetheless,
given our revised spacetime dynamics with an extended conformal gauge structure and a coupled scalar field, it seems reasonable to expect a different kinematical and perhaps more dynamical approach to the quantum black hole entropy problem. This would involve reformulating the ensembles of microstates of quantum geometry that incorporate the effects of the additional scalar field and redefine the corresponding state counting. It might also be useful to identify a quantum dissipator to allow relaxation to thermal equilibrium, ideally to achieve the corresponding Hawking temperature for black holes or Unruh temperature for accelerating frames. It would then be interesting to calculate the resulting entropies. Last but not least, given the mathematical similarity between the conformal constraint \eq{cf2} and Thiemann's complexifier\refs{Thiemann2008},
our suggested reformulation may allow to implement recently proposed new mechanism of obtaining the Bekenstein-Hawking entropy formula in LQG from an analytic continuation to $\beta=i$\refs{Frodden2014, Pranzetti2014, Achour2014, Ghosh2014} in a more natural way\refs{Refree2017}. The progress of the above continued work is deferred for future publications.

\acknowledgments

C. W. wishes to thank G. Immirzi and C. Rovelli for early discussions and brief correspondence respectively, and appreciates the EPSRC GG-Top Project and the Cruickshank Trust for financial support. O. V. is grateful to the Aberdeen University College of Physical Sciences for a research studentship.

\vspace{-5pt}

\appendix

\section{Dirac algebra of constraints in the conformal Einstein frame}
\label{app:dirac}

In this Appendix, we show that the Dirac algebra for the Hamiltonian constraint $\CC_\perp$ and diffeomorphism constraints $\CC_a$ is satisfied using the Poisson bracket $\{\cdot,\cdot\}$ with respect to the conformal Einstein frame variables $(h_{ab},p^{ab},\phi,\pi_\phi)$, where both metric and scalar fields are dynamical.

We use the Dirac $\delta$ function given by \Refs{Dew67, Han76} as a bidensity of weight zero in the first and weight one on the second argument, have the properties\refs{Dew64,Kuc95}
\begin{eqnarray}\label{eq:delta_deriv}
\delta_{,a'}(x,x')
&=&
-\delta_{,a}(x,x') ,
\ppp
\label{eq:delta_func_deriv}
f(x')\delta_{,a}(x,x')
&=&
f(x)\delta_{,a}(x,x')
%\nppp&&
+f_{,a}(x)\delta(x,x') .
%\gap
\end{eqnarray}

%\vspace{20pt}

\subsection{Poisson bracket $\left\{\CC_\perp(x),\CC_\perp(x')\right\}$}

For the Poisson bracket between two Hamiltonian constraints given by \Eq{CC2}, it is useful to consider the smeared version of the Hamiltonian constraint

\begin{equation}
\Sigma[\xi]=\int{\xi(x)\CC_\perp(x)\dd^3x},
\gap
\end{equation}
where $\xi(x)$ is an arbitrary smearing function.

%\vspace{20pt}

Applying the functional derivatives to the Hamiltonian constraints, by the antisymmetrical property of the Poisson bracket, all terms not containing any derivative of the smearing function $\xi$ and $\mu$ cancel out, and therefore what remains, after integration by parts and inserting the diffeomorphism constraint,  is
\begin{widetext} %%%%%%%%%%%%%%%%%%%%%%%%%%%%%%%%%%%%%%%%%
%\vspace{20pt}
\begin{eqnarray}\label{eq:C_C_pre_deriv}
\left\{\Sigma[\xi],\Sigma[\mu]\right\}
&=&\int\big\{-\xi_{;b}(x'')\mu(x'')h^{ab}(x'')\CC_a(x'')
+\xi(x'')\mu_{;b}(x'')h^{ab}(x'')\CC_a(x'')
\nppp&&
-\xi_{;(a}(x'')\mu(x'')\phi^{-1}(x'')\phi_{;b)}(x'')
\left[8p^{ab}(x'')-4h^{ab}(x'')p(x'')+h^{ab}(x'')\phi(x'')\pi_\phi(x'')\right]\nppp&&
+\xi(x'')\mu_{;(a}(x'')\phi^{-1}(x'')\phi_{;b)}(x'')
\left[8p^{ab}(x'')-4h^{ab}(x'')p(x'')+h^{ab}(x'')\phi(x'')\pi_\phi(x'')\right]\big\}\dd^3x''.
\gap
\end{eqnarray}
%\vspace{20pt}
\end{widetext} %%%%%%%%%%%%%%%%%%%%%%%%%%%%%%%%%%%%%%%%%
The Poisson bracket between the nonsmeared constraints is recovered by take the double functional derivative with respect to both smearing functions.
\begin{equation}
\left\{\CC_\perp(x),\CC_\perp(x')\right\}=\frac{\delta}{\delta\xi(x)}\frac{\delta}{\delta\mu(x')}
\left\{\Sigma[\xi],\Sigma[\mu]\right\}.
\gap
\end{equation}
Let us consider the second term of \Eq{eq:C_C_pre_deriv}
\begin{eqnarray*}
%\fl
F[\xi,\mu]
&=&
-\int\xi_{;(a}(x'')\mu(x'')\phi^{-1}(x'')\phi_{;b)}(x'')
\nppp&&
\times
[8p^{ab}(x'')-4h^{ab}(x'')p(x'')
\nppp&&
+h^{ab}(x'')\phi(x'')\pi_\phi(x'')]\dd^3x.
\gap
\end{eqnarray*}
Integrating by parts with respect to the first smearing function and removing the surface term, we get
\begin{eqnarray*}
\hspace{-10pt}
F[\xi,\mu]
&=&\int\xi(x)\big\{\mu(x)\phi^{-1}(x)\phi_{;(a}(x)
\times
\nppp&&\hspace{-40pt}
\left[8p^{ab}(x)-4h^{ab}(x)p(x)+h^{ab}(x)\phi(x)\pi_\phi(x)\right]\big\}_{;b)}\dd^3x.
\gap
\end{eqnarray*}
Taking the variational derivative with respect to $\xi$, the expression becomes
\begin{eqnarray*}
%\fl
\frac{\delta F}{\delta\xi(x)}&=&\int\big\{\mu(x)\phi^{-1}(x)\phi_{;(a}(x)
\,[8p^{ab}(x)
\nppp&&\hspace{-30pt}
-2h^{ab}(x)p(x)+h^{ab}(x)\left(\phi(x)\pi_\phi(x)-2p(x)\right)]\big\}_{;b)}\dd^3x.
\gap
\end{eqnarray*}
The functional derivative with respect to the second smearing function at a point $x'$, leads to
\begin{eqnarray}\label{eq:F_mu_xi}
%\fl
\frac{\delta F}{\delta\mu(x')\delta\xi(x)}
&=&
\int\big\{\phi^{-1}(x)\phi_{;(a}(x)
\nppp&&\hspace{-45pt}
\times[
8p^{ab}(x)-2h^{ab}(x)p(x)
+h^{ab}(x)(\phi(x)\pi_\phi(x)
\nppp&&\hspace{-45pt}
-2p(x))]
\delta^3(x,x')\big\}_{;b)}\dd^3x.
\gap
\end{eqnarray}
Following the same procedure on the last term of \eref{eq:C_C_pre_deriv} but taking the functional derivatives in the opposite order yields
\begin{eqnarray}\label{eq:F_xi_mu}
%\fl
\frac{\delta F'}{\delta\xi(x)\delta\mu(x')}&=&
\int\big\{\phi^{-1}(x')\phi_{;(a}(x')
\nppp&&\hspace{-45pt}
\times[
8p^{ab}(x')-2h^{ab}(x')p(x')
%\nppp&&%\hspace{-45pt}
+h^{ab}(x')(\phi(x')\pi_\phi(x')
\nppp&&\hspace{-45pt}
-2p(x'))]\delta^3(x',x)\big\}_{;b')}\dd^3x.
\gap
\end{eqnarray}

%\vspace{10pt}

By Schwarz's theorem, both derivatives can be interchanged and therefore \Eqq{eq:F_mu_xi}{eq:F_xi_mu} cancel each other out such that the Poisson bracket \eref{eq:C_C_pre_deriv} reads
\begin{eqnarray}
%\fl
\left\{\CC_\perp(x),\CC_\perp(x')\right\}&=&
\frac{\delta}{\delta\xi(x)}\frac{\delta}{\delta\mu(x')}
\nppp&&\hspace{-30pt}
\int\big\{-\xi_{;b}(x'')\mu(x'')h^{ab}(x'')\CC_a(x'')
\nppp&&\hspace{-30pt}
+\xi(x'')\mu_{;b}(x'')h^{ab}(x'')\CC_a(x'')\big\}\dd^3x''.
\gap
\end{eqnarray}
Finally, the functional derivatives are applied on the last two terms to obtain the looked for Poisson bracket between two Hamiltonian constraints
\begin{eqnarray}\label{eq:pb_CC}
\left\{\CC_\perp(x),\CC_\perp(x')\right\}
&=&
h^{ab}(x)\CC_a(x)\delta^3_{,b}(x',x)
\nppp&&
-h^{ab}(x')\CC_a(x')\delta^3_{,b'}(x,x').
\gap
\end{eqnarray}

%%%%%%%%%%%%%%%%%
\begin{widetext}

\subsection{Poisson bracket $\left\{\CC_a(x),\CC_b(x')\right\}$}

Between two diffeomorphism constraints  given by \Eq{CCi2}, the Poisson bracket is also solved using smeared constraints of the form
\begin{equation}
\Sigma[\xi^a]=\int{\xi^c(x)\CC_c(x)\dd^3x}.
\gap
\end{equation}
Here $\xi^a(x)$ is a  function for the smeared diffeomorphism constraint.
Calculating the functional derivatives of the above smeared constraints leads to the expression
\begin{eqnarray*}
%\fl
\left\{\Sigma[\xi^a],\Sigma[\mu^b]\right\}&=&\int\big\{-4\xi^{(a}(x'')\mu^b_{,a}(x'')h_{bc}(x'')p^{c)d}_{;d}(x'')-2\xi^{(a}(x'')\mu^b(x'')h_{ac,b}(x'')p^{c)d}_{;d}(x'')
\nppp&&\hspace{-50pt}
+4\xi^a_{,b}(x'')\mu^{(b}(x'')h_{ac}(x'')p^{c)d}_{;d}(x'')+2\xi^a(x'')\mu^{(b}(x'')h_{bc,a}(x'')p^{c)d}_{;d}(x'')
-\xi^a_{;a}(x'')\mu^b(x'')\phi_{;b}(x'')\pi_\phi(x'')
\nppp&&\hspace{-50pt}
-\xi^a(x'')\mu^b(x'')\phi_{;b}(x'')\pi_{\phi;a}(x'')
+\xi^a(x'')\mu^b_{;b}(x'')\phi_{;a}(x'')\pi_\phi(x'')+\xi^a(x'')\mu^b(x'')\phi_{;a}(x'')\pi_{\phi;b}(x'')\big\}\dd^3x''
\gap
\end{eqnarray*}
\end{widetext}
to be used below.
To recover the sought after Poisson brackets, the functional derivatives with respect to both smearing functions $\xi^a$ and $\mu^b$ has to be taken:
\begin{equation*}
\gap
\left\{\CC_a(x),\CC_b(x')\right\}
=\frac{\delta}{\delta\xi^a(x)}\frac{\delta}{\delta\mu^b(x')}
\left\{\Sigma[\xi^c],\Sigma[\mu^c]\right\}.
\gap
\end{equation*}
Using the definition of the diffeomorphism constraint in \Eq{CCi2}, the calculation reduces to
\begin{eqnarray}\label{eq:Ca_Cb}
\gap
\left\{\CC_a(x),\CC_b(x')\right\}&=&
\CC_b(x)\delta^3_{,a}(x',x)-\CC_a(x')\delta^3_{,b'}(x,x')
\nppp&&\hspace{-70pt}
-4h_{d[a,b]}(x)p^{dc}_{;c}(x)\delta^3(x',x)
%\nppp&&
+2\phi_{;[a}(x')\pi_\phi(x')\delta^3_{,b']}(x,x')
\nppp&&\hspace{-70pt}
+2\phi_{;[a}(x)\pi_\phi(x)\delta^3_{;b]}(x',x)
%\nppp&&
+2\phi_{;[a}(x)\pi_{\phi;b]}(x)\delta^3(x,x').
\nppp
\gap
\end{eqnarray}
Following the same procedure but exchanging the indices $a$ and $b$ yields to the Poisson bracket
\begin{eqnarray}\label{eq:Cb_Ca}
\gap
\left\{\CC_b(x),\CC_a(x')\right\}&=&\CC_a(x)\delta^3_{,b}(x',x)-\CC_b(x')\delta^3_{,a'}(x,x')
\nppp&&\hspace{-70pt}
+4h_{d[a,b]}(x)p^{dc}_{;c}(x)\delta^3(x',x)
%\nppp&&
-2\phi_{;[a}(x')\pi_\phi(x')\delta^3_{,b']}(x,x')
\nppp&&\hspace{-70pt}
-2\phi_{;[a}(x)\pi_\phi(x)\delta^3_{;b]}(x',x)
%\nppp&&
-2\phi_{;[a}(x)\pi_{\phi;b]}(x)\delta^3(x,x').
\nppp
\gap
\end{eqnarray}
Taking half of the sum of \Eqq{eq:Ca_Cb}{eq:Cb_Ca}, we obtain
\begin{equation}
\left\{\CC_{(a}(x),\CC_{b)}(x')\right\}=\CC_{(a}(x)\delta^3_{,b)}(x',x)-\CC_{(a}(x')\delta^3_{,b')}(x,x').
\gap
\end{equation}
We see that the Poisson bracket is symmetric under the exchange of the two indices; therefore, all the antisymmetric terms are equal to zero and so we have
\begin{equation*}
\left\{\CC_a(x),\CC_b(x')\right\}=
\CC_b(x)\delta^3_{,a}(x',x)-\CC_a(x')\delta^3_{,b'}(x,x').
\gap
\end{equation*}

Finally, the Poisson bracket between two diffeomorphism constraints reads
\begin{equation}\label{eq:pb_CaCb}
\left\{\CC_a(x),\CC_b(x')\right\}=\CC_a(x')\delta^3_{,b'}(x,x')+\CC_b(x)\delta^3_{,a}(x',x).
\gap
\end{equation}

%\vspace{5pt}

\subsection{Poisson bracket $\left\{\CC_a(x),\CC_\perp(x')\right\}$}

To derive the Hamiltonian-diffeomorphism Poisson bracket, we use the fact that the Poisson bracket between the diffeomorphism constraint and any weight-one element $f$, as e.g. the Hamiltonian constraint, gives the same result:
\begin{equation}\label{eq:PB_weight_1}
\left\{\CC_c(x),f(x')\right\}=f(x)\delta^3_c(x,x').
\gap
\end{equation}
Using its linearity, we can obtain the Poisson bracket by calculating it for every term of the Hamiltonian constraint separately. Considering the first term of \Eq{CC2},
\begin{equation}
S=\frac{1}{\sqrt{h}}\phi^{-2}p_{ab}p^{ab}
\gap
\end{equation}
and using the properties of the Dirac $\delta$ function \eref{eq:delta_deriv} and \eref{eq:delta_func_deriv}, the first Poisson bracket can be calculated. It is derived by separating it into simpler terms using the property
\begin{equation}
\left\{f,gh\right\}=\left\{f,g\right\}h+g\left\{f,h\right\}.
\gap
\end{equation}
\begin{widetext}
We therefore obtain using the variation of $h^{-1/2}$ that

\begin{eqnarray*}\gap
%\fl
\left\{\CC_e(x),S(x')\right\}
&=&
\Big\{\CC_e(x),\frac{2\kappa}{\sqrt{h(x')}}\Big\}
\phi^{-2}(x')h_{a(c}(x')h_{d)b}(x')p^{ab}(x')p^{cd}(x')
\nppp&&\hspace{-65pt}
+\frac{2\kappa}{\sqrt{h(x')}}\left\{\CC_e(x),\phi^{-2}(x')\right\}h_{a(c}(x')h_{d)b}(x')p^{ab}(x')p^{cd}(x')
%\nppp&&
+\frac{2\kappa}{\sqrt{h(x')}}\phi^{-2}(x')\left\{\CC_e(x),h_{a(c}(x')\right\}h_{d)b}(x')p^{ab}(x')p^{cd}(x')
\nppp&&\hspace{-65pt}
+\frac{2\kappa}{\sqrt{h(x')}}\phi^{-2}(x')h_{a(c}(x')\left\{\CC_e(x),h_{d)b}(x')\right\}p^{ab}(x')p^{cd}(x')
%\nppp&&
+\frac{4\kappa}{\sqrt{h(x')}}\phi^{-2}(x')p_{ab}(x')\left\{\CC_e(x),p^{ab}(x')\right\}.
\gap
\end{eqnarray*}
\end{widetext}
This expression after some calculations reduces to the Poisson bracket
\begin{equation}
\left\{\CC_a(x),S(x')\right\}=S(x)\delta_{,a}(x,x').
\gap
\end{equation}
This shows that the first term follows the weight-one element rule \eref{eq:PB_weight_1}. By extending linearly to the other terms, we extrapolate to the global Hamiltonian constraint and therefore have
\begin{equation}\label{eq:pb_CeC}
\left\{\CC_a(x),\CC_\perp(x')\right\}=\CC_\perp(x)\delta_{,a}(x,x').
\gap
\end{equation}

Therefore, combining all results in this Appendix, we have the following Poisson brackets \Eqqq{eq:pb_CC}{eq:pb_CaCb}{eq:pb_CeC}:
%\numparts
\begin{eqnarray}
\label{eq:PB-CC}
\left\{\CC_\perp(x),\CC_\perp(x')\right\}
&=&
h^{ab}(x)\CC_a(x)\delta^3_{,b}(x',x)
\nppp&&%\hspace{20pt}
-h^{ab}(x')\CC_a(x')\delta^3_{,b'}(x,x')
\\\label{eq:PB-CaCb}
\left\{\CC_a(x),\CC_b(x')\right\}
&=&
\CC_a(x')\delta^3_{,b'}(x,x')
\nppp&&%\hspace{20pt}
+\CC_b(x)\delta^3_{,a}(x',x)
\\\label{eq:PB-CeC}
\left\{\CC_e(x),\CC_\perp(x')\right\}
&=&
\CC_\perp(x)\delta^3_{,e}(x,x')
\gap
\end{eqnarray}
which all consistently vanish weakly if the constraints $\CC_\perp(x)$ and $\CC_a(x)$ vanish weakly.
The above relations form the same Dirac algebra of constraints as with the metric tensor-only theory of GR\refs{Bro95}.

%%%%%%%%%%%%%%%%%%%{\color{gray}\subsection*{Argument about why only for these variables}}%%%%%%%%%%%%%%%%%%%%%%

Furthermore, since the conformal Ashtekar-Barbero variables are constructed from the conformal Einstein frame ST variables using a set of canonical transformations, the argument about the previous Poisson brackets is also valid\refs{Boj15, Boj10} in these new variables in \Eq{cano1} or \Eq{cano2}.

%\newpage
\section{Global conformal transformation relations}
\label{apx:conf}

Here we summarize how various physical quantities used in this work undergo changes with a global conformal transformation used in the main text of this work.
It follows directly from \Eq{cnf1} that, under a global conformal transformation given by \Eq{Omg}, we have the following relations
\begin{eqnarray}\gap
h_{ab}
&\rightarrow&
\Lambda^2 h_{ab}
\label{confh}
\ppp
h^{ab}
&\rightarrow&
\Lambda^{-2} h^{ab}
\ppp
\sqrt{h}
&\rightarrow&
\Lambda^3 \sqrt{h}
\ppp
N
&\rightarrow&
\Lambda N
\ppp
N^a
&\rightarrow&
N^a
\ppp
N_a
&\rightarrow&
\Lambda^2 N_a .
\gap
\end{eqnarray}
%
%\newpage\noindent
Using the above and \Eq{defKij} we have
\begin{eqnarray}\gap
K_{ab}
&\rightarrow&
\Lambda K_{ab}
\ppp
K^{ab}
&\rightarrow&
\Lambda^{-3} K^{ab}
\label{Kcnf2}
\ppp
K
&\rightarrow&
\Lambda^{-1} K .
\label{Kcnf3}
\gap
\end{eqnarray}
%
%\newpage\noindent
Furthermore, from \Eqq{eq43}{eq45} we have
\begin{eqnarray}\gap
p^{ab}
&\rightarrow&
\Lambda^{-2}
p^{ab}
\label{eq43cnf1}
\ppp
p_{ab}
&\rightarrow&
\Lambda^{2}
p_{ab}
\label{eq43cnf2}
\ppp
p
&\rightarrow&
p
\label{eq43cnf3}
\ppp
\pi_\phi
&\rightarrow&
\Lambda\pi_\phi
\label{eq45cnf}
\ppp
R[h]
&\rightarrow&
\Lambda^{-2} R[h] .
\label{eq46cnf}
\gap
\end{eqnarray}
From \Eqqq{CC2}{CCi2}{cf} we see that
\begin{eqnarray}\gap
\CC_\perp
&\rightarrow&
\Lambda^{-1} \CC_\perp
\label{CCcnf}
\ppp
\CC_a
&\rightarrow&
\CC_a
\label{CCicnf}
\ppp
\CC
&\rightarrow&
\CC
\label{CCCcnf}
\gap
\end{eqnarray}
and hence
\begin{eqnarray}
\LL_\text{ES}
&\rightarrow&
\LL_\text{ES}
\label{LLcnf}
\\
\HH_\text{ES}
&\rightarrow&
\HH_\text{ES} .
\label{HHcnf}
\gap
\end{eqnarray}
In terms of the triad variables, we find that

\begin{eqnarray}
e^i_a
&\rightarrow&
\Lambda e^i_a
\label{ecnf1}
\ppp
e^a_i
&\rightarrow&
\Lambda^{-1} e^a_i
\label{ecnf2}
\gap
\end{eqnarray}

\begin{eqnarray}
E^i_a
&\rightarrow&
\Lambda^{-2} E^i_a
\label{Ecnf1}
\ppp
E^a_i
&\rightarrow&
\Lambda^2 E^a_i
\label{Ecnf}
\ppp
K^i_a
&\rightarrow&
\Lambda^{-2} K^i_a .
\label{Kcnf}
\gap
\end{eqnarray}

Finally, from \Eqqq{spincons}{Ecnf}{Kcnf} we have
\begin{eqnarray}
\CC_k
&\rightarrow&
\CC_k .
\label{CCkcnf}
\gap
\end{eqnarray}

\end{document}